\documentclass[aps,prd,twocolumn,superscriptaddress,nofootinbib,amsmath,amssymb]{revtex4-2}
\usepackage{graphicx}
\usepackage{bm}
\usepackage{xcolor}
\usepackage[colorlinks=true,citecolor=blue,urlcolor=blue,linkcolor=red]{hyperref}
\definecolor{hunter}{RGB}{53,94,59}      
\allowdisplaybreaks

\begin{document}

\title{Wormholes in Vegh--dRGT massive gravity: a null-energy transparency theorem, horizon obstruction, and exact isotropic solutions}

\author{Sudan Hansraj}
\email[Corresponding author: ]{hansrajs@ukzn.ac.za}
\author{Ayan Banerjee} 
\affiliation{Astrophysics Research Centre, School of Mathematics, Statistics
and Computer Science, University of KwaZulu-Natal, Private Bag X54001,
Durban 4000, South Africa}

\date{\today}

\begin{abstract}
We study static, spherically symmetric wormholes in de
Rham--Gabadadze--Tolley massive gravity with the spatially degenerate Vegh
reference metric. The effective fluid carried by the massive graviton
contributes nothing to the combination of density and radial pressure that the
flare-out condition constrains, so it is invisible to the null energy condition
along radial light rays. This holds for the whole degenerate class of reference
metrics, for all four interaction potentials, and for any static spherically
symmetric geometry; the same property singles out the Vegh choice as the only
consistent one. The matter alone must therefore violate the null energy
condition at every throat, exactly as in general relativity. This corrects a
claim repeated across the massive-gravity wormhole literature. We then determine
where throats may sit. Wormholes with isotropic pressure fall into two families.
One lies beyond the cosmological horizon of the theory's own vacuum and has no
static exterior; it is empty in general relativity and is switched on by the
graviton coupling. The other lies inside that horizon and needs a steeply
falling redshift. We give exact perfect-fluid solutions for both. In the second
family the graviton term can turn the tension at the throat into ordinary
pressure. A thin shell closes the configuration, and its surface stresses and
its exotic-matter content follow in closed form. The shells are radially
unstable, and light rays near the throat trace an unstable circular orbit.
\end{abstract}

\maketitle

\section{Introduction}
\label{sec:intro}

A wormhole is a tunnel in spacetime. It links two regions through a
narrow neck, called the throat, where the areal radius is smallest. The idea
began with the bridge solution of Einstein and Rosen \cite{Einstein:1935particle}
and with the geometrodynamics of Misner and Wheeler \cite{Misner:1957classical}.
Morris and Thorne \cite{morris} made it a concrete problem. They asked what
matter is needed to hold the throat open. Their answer is the flare-out
condition, and it comes at a cost: the null energy condition (NEC) must fail at
the throat. The same conclusion follows from the causality study of Morris,
Thorne and Yurtsever \cite{Morris:1988wormholesc} and from the cut-and-paste
constructions of Visser \cite{Visser:1989traversable}. The result is standard
and is set out in the book of Visser \cite{visserbook}, in the throat analysis
of Hochberg and Visser \cite{hochberg}, and in the review of Lobo
\cite{Lobo:2017wormhole}. Matter that breaks the NEC is called exotic. No known
laboratory fluid behaves in this way.

Since the violation cannot be removed, the next question is how much of
it is needed. Visser, Kar and Dadhich \cite{vkd} answered this with a volume
integral of $\rho+p_r$, which measures the total exotic content of a
configuration. Nandi, Zhang and Kumar \cite{Nandi:2004volume} later sharpened
the measure. Both show that the exotic content can be made as small as one
likes. One cuts the wormhole off close to the throat and glues on a normal
exterior across a thin shell. This is the surgical method of Visser
\cite{Visser:1989traversablea}, developed further for thin shells in general
dimension by Dias and Lemos \cite{Dias:2010thinshell}.

One line of work keeps general relativity (GR) and looks for unusual
matter. Phantom energy, which has $p<-\rho$, breaks the NEC by construction.
Sushkov \cite{Sushkov:2005wormholes} and Lobo
\cite{Lobo:2005phantom,Lobo:2005stability} built wormholes from it. Chaplygin
gas and its generalisations were used in the same way
\cite{Lobo:2006chaplygin,Kuhfittig:2009single}, and Gonz\'alez et al.\
\cite{Gonzalez:2009wormholes} studied how such throats evolve. In all of these
the exotic burden stays with the matter.

A second line of work changes the gravity instead. The extra terms in the
field equations are moved to the right-hand side and read as an effective
fluid. The hope is that this fluid, and not the real matter, breaks the NEC at
the throat. The $f(R,T)$ theory of Harko et al.\ \cite{Harko:2011gravity} has
been used in this way many times
\cite{Zubair:2016static,Moraes:2017modeling,Elizalde:2018wormhole,Sahoo:2018phantom,Elizalde:2019wormholes,Godani:2019static,Moraes:2019charged,Banerjee:2020conformally,Mishra:2020traversable,Chanda:2021morris,Rastgoo:2025wormholes},
and some of these papers report throats held open by matter that is not exotic
\cite{Zubair:2019exact,LuisRosa:2022nonexotica}. The junction conditions of the
theory have been worked out separately \cite{Rosa:2021junction}. Rastall gravity
has been used for the same purpose
\cite{Moradpour:2017traversable,Halder:2019wormhole,Mustafa:2020stable,Mustafa:2020noncommutative}.
Every such claim rests on one point. The effective fluid must contribute to the
sum $\rho+p_r$ at the throat. If it does not, the claim cannot hold, however the
rest of the calculation is arranged. We settle this point for massive gravity.

Massive gravity gives the graviton a mass. The linear theory carries a
ghost. The cure was found by de Rham and Gabadadze
\cite{Rham:2010generalization} and completed by de Rham, Gabadadze and Tolley
\cite{dRGT}. Hassan and Rosen \cite{hassanrosen} then proved that the theory is
free of the ghost at full nonlinear order. The construction is reviewed in
Refs.~\cite{Hinterbichler:2012theoretical,Rham:2014massive}. The mass term is
built from a fixed reference metric, which does not obey field equations of its
own. Its contribution can also be read as an effective fluid. That fluid is
often called dark-energy-like, and this description has been taken to mean that
it breaks the energy conditions and can therefore hold a throat open.

The reference metric has to be chosen. We use the spatially degenerate
choice of Vegh \cite{vegh}, in which the time and radial entries vanish and only
the angular block survives. It was introduced in a holographic setting, where it
relaxes momentum \cite{Davison:2013momentum} and where the graviton mass comes
from a lattice \cite{Blake:2014holographic}. It is also the choice used in the
black-hole solutions of Cai et al.\ \cite{cai} and in later work on their thermodynamics
\cite{gtw}. Compact objects have been studied in dRGT
gravity as well: relativistic stars \cite{Katsuragawa:2016relativistic}, neutron
stars \cite{Hendi:2017neutron} and white dwarfs \cite{Panah:2019white}. The stellar-structure problem in this gauge is treated separately by the present authors. The wormhole problem is set here in the same canonical variables.

The wormhole literature in dRGT gravity is short, and it is uniform in one
respect. Tangphati et al.\ \cite{tangphati} studied traversable wormholes in
$f(R)$ massive gravity with an exponential shape function put in by hand, so the
imprint of the graviton on the geometry was never extracted. Kamma et al.\
\cite{kamma} treated wormholes on a massive-graviton background in a short
proceedings contribution. The fullest treatment is that of Dutta, Roy,
Pullisseri and Chakraborty \cite{dutta}. They built ultrastatic anisotropic
wormholes in dRGT--$f(R,T)$ and in dRGT--Einstein gravity, using the
parameterisation of Ghosh, Tannukij and Wongjun \cite{gtw}, in which the
graviton sector carries the effective stresses
\begin{equation}
\rho^{(g)} = -\frac{2\tilde\gamma-\Lambda r}{8\pi r},\quad
p_r^{(g)} = +\frac{2\tilde\gamma-\Lambda r}{8\pi r},\quad
p_t^{(g)} = \frac{\tilde\gamma-\Lambda r}{8\pi r}.
\label{duttastress}
\end{equation}
Bhar \cite{Bhar:2024properties} produced two further wormhole families in
the same theory, Kumar, Maurya and Kiroriwal \cite{Kumar:2024dark} fixed the
shape function from dark-matter density profiles, and Chaudhary, Maurya and
Kumar \cite{Chaudhary:2025wormhole} did the same using galactic rotation curves.
Evolving wormholes were treated in Ref.~\cite{evolving}, and Landry et al.\
\cite{landry} studied Morris--Thorne geometries in $F(T)$ teleparallel gravity
with a dRGT mass term added. Across this body of work the same statement recurs:
the graviton stresses violate the energy conditions, and so a wormhole in
massive gravity is a natural outcome.

Two things motivate the present paper. The first is a gap. No study has
treated the spatially degenerate Vegh gauge. Its effective fluid is a constant
shift $\alpha-1=m^2C^2c_2$ together with a linear term of coefficient
$\gamma/2=m^2c_1C/2$. This differs in structure from (\ref{duttastress}), and it is the sector treated here for the first time in the wormhole context. The second is an error in the standing
interpretation. The stresses (\ref{duttastress}) satisfy
$\rho^{(g)}+p_r^{(g)}=0$ on inspection, so the graviton sector adds nothing to
the sum that the flare-out condition constrains. Yet Ref.~\cite{dutta} concludes
that the massive gravitons supply the negative energy density that keeps the
throat open, and reports ordinary matter with $\rho+p_r\ge 0$ near the throat of
an Einstein--massive wormhole. We show below that both statements conflict with
the flare-out identity, and that they do so under the effective-stress split
displayed in Ref.~\cite{dutta} itself.

Our first result is that this is not an isolated slip. The property
$\rho^{(g)}+p_r^{(g)}=0$ holds for the whole diagonal spatially degenerate class
of reference metrics, for all four dRGT potentials, and for an arbitrary static
spherically symmetric metric. It is already visible in (\ref{duttastress}). It
forces the matter alone to carry the NEC violation at every throat. The
transparency theorem therefore applies to each of the constructions listed
above, and any relief of the matter sector reported in them must come from the
curvature or torsion terms, never from the mass term.

The plan and the main results are as follows. Section~\ref{sec:framework}
sets out the field equations and the canonical variables and translates the Morris--Thorne conditions into them: a
throat is a simple zero of the spatial potential. Section~\ref{sec:transp}
proves the transparency theorem and draws its corollaries.
Section~\ref{sec:dichotomy} derives the throat-slope formula for isotropic
pressure and establishes a dichotomy between super-horizon wormholes with
non-decreasing redshift and sub-horizon wormholes with steeply decreasing
redshift; the ultrastatic isotropic class, empty in GR, exists in
Vegh--dRGT gravity only beyond the cosmological zero of its own vacuum, and the
coupling bound quoted in Sec.~\ref{sec:dichotomy} pushes such throats beyond
$\sim\!50$ km. Sections~\ref{sec:ultra} and \ref{sec:sub} exhibit
exact perfect-fluid wormholes realising each branch, the second of which
possesses a universal throat density and a graviton-generated window of
positive throat pressure. Section~\ref{sec:shell} completes the sub-horizon
configuration with a Darmois--Israel shell joining it to the exact vacuum and evaluates the surface stresses and the exotic budget in
closed form. Section~\ref{sec:deflection} develops a finite-radius
light-propagation diagnostic for the static-patch completion, and
Section~\ref{sec:disc} summarises and outlines the programme.
All symbolic claims in this paper have been verified by computer algebra.

\section{Field equations, canonical variables, and the throat}
\label{sec:framework}

We work with the dRGT action in four dimensions with couplings $c_1,c_2$
and the degenerate Vegh reference metric
$f_{ab} = \mathrm{diag}(0,0,C^2,C^2\sin^2\theta)$, where $C$ carries the
dimension of length \cite{vegh,cai}. For the static, spherically
symmetric line element
\begin{equation}
ds^2 = -e^{2\nu(r)}dt^2 + e^{2\lambda(r)}dr^2 + r^2 d\Omega^2 ,
\label{metric}
\end{equation}
the field equations of an anisotropic fluid read
\begin{align}
\kappa r^{2}\rho &= e^{-2\lambda}(2r\lambda'-1) + \alpha + \gamma r,
\label{fe1}\\
\kappa r^{2}p_{r} &= e^{-2\lambda}(2r\nu'+1) - \alpha - \gamma r,
\label{fe2}\\
\kappa r^{2}p_{t} &= e^{-2\lambda}\bigl(r\nu'+r^2\nu'^2-r\lambda'
-r^2\lambda'\nu'+r^2\nu''\bigr) - \tfrac{1}{2}\gamma r,
\label{fe3}
\end{align}
with $\alpha = 1+m^2C^2c_2$ and $\gamma = m^2c_1C$, and the unique static vacuum is $e^{-2\lambda} = e^{2\nu} = \alpha + \tfrac{1}{2}\gamma r - 2M/r$. In the canonical variables
\begin{equation}
x = Kr^2,\qquad e^{-2\lambda} = Z(x),\qquad e^{2\nu} = y^2(x),
\end{equation}
the condition of isotropic pressure $p_r=p_t=p$ is the master equation
\begin{equation}
4x^2 Z\ddot y + 2x^2\dot Z\dot y
+ \bigl(\dot Z x - Z + \alpha + \beta\sqrt{x}\bigr)y = 0,
\label{master}
\end{equation}
with $\beta = \gamma/(2\sqrt{K})$, and the density and pressure are
\begin{equation}
\frac{\kappa\rho}{K} = -2\dot Z - \frac{Z-\alpha}{x} + \frac{2\beta}{\sqrt x},
\qquad
\frac{\kappa p}{K} = \frac{4Z\dot y}{y} + \frac{Z-\alpha}{x}
- \frac{2\beta}{\sqrt x}.
\label{thermo}
\end{equation}
For a prescribed seed $y(x)$, (\ref{master}) is linear and first order in
$Z$,
\begin{equation}
(2x^2\dot y + xy)\dot Z + (4x^2\ddot y - y)Z + (\alpha+\beta\sqrt{x})y = 0,
\label{linearZ}
\end{equation}
so that every tractable seed generates an exact solution by quadrature.

The Morris--Thorne form
$ds^2 = -e^{2\Phi}dt^2 + (1-b/r)^{-1}dr^2 + r^2d\Omega^2$ corresponds to
$Z = 1-b/r$ and $y = e^{\Phi}$. A throat $b(r_0)=r_0$ with flare-out
$b'(r_0)<1$ and no horizon translates into three statements about the
canonical potentials: $Z(x_0)=0$ is a \emph{simple zero} approached from a
region $Z>0$; the slope is positive, $\dot Z(x_0)>0$, since
$Z'(r) = (b-rb')/r^2 = (1-b')/r_0$ at the throat; and $y(x_0)$ is finite and
nonzero. The proper radial distance $\int dr/\sqrt{Z}$ converges across a
simple zero, the areal radius is minimised there, and the geometry doubles
across the throat in the usual way. The centre of the stellar problem is
thus replaced, as the distinguished locus, by a zero of $Z$, and the same canonical apparatus applies.

\subsection{The two potentials and the geometry of their zeros}
\label{ssec:model}

There are two metric functions that enter the construction below and must be kept distinct. The first is the interior spatial potential $Z(x)=e^{-2\lambda}$, which together with the independent redshift $y^{2}(x)=e^{2\nu}$ governs the matter-filled region containing the throat; the two are tied to one another only through the master equation
(\ref{master}). The second is the vacuum potential
\begin{equation}
Z_{\rm vac}(x)=\alpha+\beta\sqrt{x}-\frac{2M\sqrt{K}}{\sqrt{x}},
\label{Zvac}
\end{equation}
the unique static exterior of the theory, which contains no matter and in which the two potentials collapse onto one another, $e^{2\nu}=e^{-2\lambda}=Z_{\rm vac}$. It is of Schwarzschild--de Sitter type, with $\alpha$ in place of unity and with a graviton term linear, rather than quadratic, in the areal radius. The functions $Z$ and
$Z_{\rm vac}$ are never simultaneously active: on the interior interval the geometry is $Z$, on the exterior interval it is $Z_{\rm vac}$, and the two are related only by the Darmois--Israel conditions imposed at a single 
matching radius $x_{b}$, which fix the shell stresses and the mass $M$. They do not coincide even there --- the shell exists precisely because $Z$, $Z_{\rm vac}$ and their slopes differ at $x_{b}$.

This distinction determines the character of the zeros, and three geometrically different situations arise. A simple zero of $Z$ approached from $Z>0$ with $\dot Z(x_{0})>0$ is a \emph{throat}: the proper radial distance $\int dx/(2\sqrt{Kx}\,\sqrt{Z})$ converges across it, the areal radius is minimal, and the geometry flares outward. It is never a horizon, because the zero sits in the \emph{spatial} potential, so that $g_{rr}=1/Z$ diverges while $g_{tt}=-y^{2}$ stays finite and nonzero. A 
simple zero of $Z$ with $\dot Z<0$ is the opposite object: it is a sphere of \emph{maximal} area at which the static Lorentzian region terminates, $Z$ becoming negative beyond it. It is not a throat, however exactly the condition $b(r_{0})=r_{0}$ is satisfied there. A zero of $Z_{\rm vac}$, by contrast, is always a Killing horizon, because the static vacuum has $g_{tt}=-Z_{\rm vac}$, so that the vanishing of the potential is the vanishing of $g_{tt}$ and $\partial_{t}$ becomes null; the outer such zero
$x_{c}=Kr_{c}^{2}$, fixed by $\alpha^{2}\geq4|\gamma|M$ as in (\ref{rc}) below, is the cosmological horizon of the theory's own vacuum and the outer boundary of its static patch, entirely analogous to the horizon of the de~Sitter static patch. The matching radius $x_{b}$ is a zero of
\emph{neither} function; it is a free parameter of the construction. The whole physical content of the model therefore requires only that the window
\begin{equation}
x_{0} \;<\; x_{b} \;<\; x_{c}
\label{ordering}
\end{equation}
be non-empty --- equivalently, that the throat lie inside the static patch of its own vacuum. Whether this can be organised is determined by the field equations, and that is the subject of the dichotomy of  Sec.~\ref{sec:dichotomy}.

\section{Radial null-energy transparency of the graviton sector}
\label{sec:transp}

Consider the general diagonal spatially degenerate reference metric
\begin{equation}
f_{ab}=\mathrm{diag}\bigl(0,\,0,\,F(r),\,F(r)\sin^2\theta\bigr),
\qquad F(r)>0,
\label{fgen}
\end{equation}
of which the Vegh gauge is the constant profile $F=C^2$, inserted into the
\emph{arbitrary} static metric (\ref{metric}) --- no field equations are
used. The matrix $K^{a}{}_{b}=\sqrt{g^{ac}f_{cb}}$ is diagonal with exactly
two equal nonvanishing eigenvalues $k=\sqrt{F}/r$; throughout we take the
principal square root of the positive semi-definite matrix $g^{ac}f_{cb}$,
which identifies the physical diagonal branch. With two equal eigenvalues
one finds
$\mathcal{U}_3=\mathcal{U}_4=0$ identically, the couplings $c_3$ and $c_4$
drop out of the interaction tensor altogether, and a direct computation
retaining all four potentials gives
\begin{equation}
\chi^{t}{}_{t} = \chi^{r}{}_{r} = -\frac{c_1\sqrt{F}}{r} - \frac{c_2F}{r^2},
\qquad
\chi^{\theta}{}_{\theta} = -\frac{c_1\sqrt{F}}{2r},
\label{chis}
\end{equation}
verified symbolically for arbitrary $F(r)$, $\Phi(r)$ and $b(r)$: an
algebraic property of the whole class (\ref{fgen}), for any couplings.

The generalisation is, however, disciplined by consistency. With a fixed
non-dynamical reference metric the contracted Bianchi identity and matter
conservation force $\nabla_a\chi^{a}{}_{b}=0$ on shell, and transparency
itself sharpens this condition remarkably: in the divergence
$\nabla_a\chi^{a}{}_{r} = \partial_r\chi^{r}{}_{r}
+ \nu'\,(\chi^{r}{}_{r}-\chi^{t}{}_{t})
+ (2/r)(\chi^{r}{}_{r}-\chi^{\theta}{}_{\theta})$
the metric enters only through $\nu'$, whose coefficient vanishes by
(\ref{chis}), so the consistency condition is \emph{metric-independent}
and constrains the fiducial profile alone:
\begin{equation}
\nabla_a\chi^{a}{}_{r}
= -\,\frac{\bigl(c_1 r + 2c_2\sqrt{F}\bigr)F'(r)}{2r^{2}\sqrt{F}} \;=\;0,
\label{bianchi}
\end{equation}
verified symbolically, with the $t$- and angular components vanishing
identically. For a smooth positive profile $F(r)$ and a nontrivial
interaction sector $(c_1,c_2)\neq(0,0)$, equation (\ref{bianchi}) admits
two branches: the constant profile $F=C^2$, which is the Vegh gauge; and
the fine-tuned branch $\sqrt{F}=-c_1 r/(2c_2)$ (requiring $c_1c_2<0$), on which
$k=-c_1/(2c_2)$ is constant and the interaction tensor collapses to
$\chi^{a}{}_{b}=\bigl[c_1^{2}/(4c_2)\bigr]\delta^{a}{}_{b}$ --- a pure
effective cosmological constant with no independent graviton imprint.
Within the diagonal degenerate class, therefore, the Vegh profile is not
merely a convenient choice but the \emph{unique} consistent fiducial with
nontrivial graviton stresses, and every consistent member of the class is
radially transparent. Sectors such as that of Refs.~\cite{gtw,dutta},
whose effective stresses arise from genuine St\"uckelberg dynamics rather
than a fixed diagonal fiducial, restore conservation through the scalar
equations of motion; for them transparency is confirmed by direct
inspection of the displayed stresses (\ref{duttastress}), which satisfy
$\rho^{(g)}+p_r^{(g)}=0$ manifestly. For the remainder of the paper we
work in the Vegh gauge, $F=C^2$, with $C$ carrying the dimension of
length. Writing the
graviton contribution as an effective fluid,
$\kappa\rho^{(g)} = m^2\chi^{t}{}_{t}$ and
$\kappa p_r^{(g)} = -m^2\chi^{r}{}_{r}$ up to the sign conventions of
(\ref{fe1})--(\ref{fe2}), the first equality in (\ref{chis}) is the
statement
\begin{equation}
\rho^{(g)} + p_r^{(g)} = 0 \qquad \text{pointwise, for any static metric.}
\label{transparency}
\end{equation}
We refer to (\ref{transparency}) as \emph{radial null-energy transparency}:
along radial null rays the graviton sector is invisible to the Raychaudhuri
focusing that the flare-out condition encodes. By the classification above
it is shared by every consistent member of the diagonal degenerate class,
and it holds by inspection in the St\"uckelberg-based sector
(\ref{duttastress}) as well: no choice of couplings $c_1,\dots,c_4$ or of
angular profile evades it.

The corollary is immediate and sharp. Adding (\ref{fe1}) and (\ref{fe2})
for the Morris--Thorne metric, the graviton terms cancel and one obtains,
at a throat where $b(r_0)=r_0$ and $\Phi$ is finite,
\begin{equation}
\kappa\bigl(\rho+p_r\bigr)\Big|_{r_0} = \frac{b'(r_0)-1}{r_0^{2}} \;<\; 0,
\label{necthroat}
\end{equation}
with no dependence on $m$, $c_1$, $c_2$ or $C$ whatsoever. The matter
sector alone must violate the radial NEC at the throat of every
Morris--Thorne wormhole in Vegh--dRGT massive gravity, exactly as in GR.
The graviton sector can neither supply this violation nor be arranged to
evade it. In particular, under the effective-stress decomposition displayed
in Ref.~\cite{dutta}, whose stresses (\ref{duttastress}) satisfy
(\ref{transparency}) manifestly, the Einstein--massive claim that ordinary
matter can satisfy the radial null energy condition at the throat while the
gravitons supply the required negative energy is incompatible with the
flare-out identity (\ref{necthroat}). In the
$f(R,T)$-massive and $F(T)$-massive constructions
\cite{dutta,landry}, any genuine relief of the matter sector must originate
in the curvature or torsion terms, never in the mass term.

This can be made quantitative against the published numerical calculations. This is important because Fig.~5 of Ref.~\cite{dutta} seems to show throats at which the null energy condition is satisfied. We adopt for this paragraph the notation of that work, writing 
$\alpha_{\rm f},\beta_{\rm f}$ for its $f(R,T)$ coefficients and $\gamma,\Lambda$ for its graviton parameters, which are not those of (\ref{fe1})--(\ref{fe3}). For the linear choice $f=\alpha_{\rm f}R+\beta_{\rm f}T$, so that $f_{R}$ is constant, the matter density and radial pressure are its Eqs.~(29) and (30), and their sum is 
\begin{equation}
\rho+p_{r}=\frac{\alpha_{\rm f}\,\bigl(rb'-b\bigr)}
{r^{3}\,\bigl(8\pi+\beta_{\rm f}\bigr)},
\label{duttasum}
\end{equation}
the graviton terms in $\gamma$ and $\Lambda$ having cancelled identically: these are already visible in the equations of Ref.~\cite{dutta} itself. Since flare-out is the condition $b-rb'>0$ at $r_{0}$, the right-hand side of (\ref{duttasum}) is negative at any genuine throat, for every value of the graviton parameters. The resolution of the apparent conflict therefore lies in the shape function.
Differentiating Eq.~(36) of Ref.~\cite{dutta} at the throat, for the parameters of its Fig.~5 ($\alpha_{\rm f}=1$, $\beta_{\rm f}=2$, $\sigma=-1$), gives the closed form 
\begin{equation}
b'(r_{0}) = -1 + \frac{4(4\pi+1)}{4\pi+2}\,
r_{0}\bigl(\Lambda r_{0}-2\gamma\bigr),
\label{duttaslope}
\end{equation}
so that flare-out at $r_{0}=1$ requires
$\Lambda-2\gamma<(4\pi+2)/[2(4\pi+1)]\simeq0.537$. Evaluating
(\ref{duttaslope}) on the four parameter sets of that figure, all at
$r_{0}=1$, yields
\begin{equation}
b'(r_{0})=\begin{cases}
+1.533, & \gamma=0.01,\ \Lambda=0.7,\\[2pt]
+1.682, & \gamma=-0.01,\ \Lambda=0.7,\\[2pt]
-3.682, & \gamma=0.01,\ \Lambda=-0.7,\\[2pt]
-1.004, & \gamma=0.001,\ \Lambda=0.001.
\end{cases}
\label{duttabp}
\end{equation}
The first two exceed unity. For them flare-out fails at $r=r_{0}$, where
$Z=1-b/r$ has a simple zero of \emph{negative} slope: by the
classification of Sec.~\ref{ssec:model} that surface is a maximal-area
sphere terminating the static region, not a throat, and the assumed
domain $r>r_{0}$ is not static and Lorentzian. These are precisely the
two cases for which Table~I of Ref.~\cite{dutta} reports $\rho+p_{r}\ge0$
from $r\simeq0.87$ and $r\simeq0.83$ outward. The remaining two sets do
satisfy flare-out, and for both the same table records $\rho+p_{r}<0$ at
$r_{0}=1$ --- negative everywhere in the one case, and out to
$r\simeq6.09$ in the other. With this understanding the published figure is not a counterexample to transparency but an illustration of it. We find that every configuration in it that flares out violates the radial null energy condition at its throat, and every  configuration that does not violate it 
is not a wormhole. The mechanism is moreover recognisably that of Sec.~\ref{sec:dichotomy}. An ultrastatic redshift together with a sufficiently large positive effective cosmological term turns the surface $b(r_{0})=r_{0}$ from a minimal into a maximal sphere; the dichotomy
established below is the isotropic form of that statement, and (\ref{duttaslope}) is its anisotropic counterpart in the variables of Ref.~\cite{dutta}.

Transparency is radial, not total. With the sign conventions of
(\ref{fe1})--(\ref{fe3}), $\kappa\rho^{(g)} = (1-\alpha)/r^2 - \gamma/r$
and $\kappa p_t^{(g)} = \gamma/(2r)$, so the tangential combination is
\begin{equation}
\rho^{(g)} + p_t^{(g)} = \frac{1}{\kappa}\left(\frac{1-\alpha}{r^2}
- \frac{\gamma}{2r}\right),
\end{equation}
which is non-negative for the physical sign $\gamma<0$ (taking $\alpha\leq 1$): the graviton
sector violates neither the radial nor, for these signs, the tangential
null energy condition. It nonetheless participates fully in the tangential
stress balance, reshapes the geometry away from the throat, and thereby
constrains both where throats can sit and how the exotic matter is
distributed. The remainder of this paper quantifies exactly that.

\section{The throat-slope formula and a horizon dichotomy}
\label{sec:dichotomy}

For isotropic pressure the master equation governs everything, and
evaluating (\ref{linearZ}) on a throat $Z(x_0)=0$ yields the slope in
closed form,
\begin{equation}
\dot Z(x_0) \;=\; -\,
\frac{\bigl(\alpha+\beta\sqrt{x_0}\bigr)\,y(x_0)}
{x_0\bigl[\,2x_0\dot y(x_0)+y(x_0)\,\bigr]}\,.
\label{slope}
\end{equation}
Equation (\ref{slope}) is the isotropic wormhole existence test: a seed
$y(x)$ admits a throat at $x_0$ if and only if the right-hand side is
positive there.

Consider first any redshift function that is non-decreasing at the throat,
$\dot y(x_0)\geq 0$ with $y>0$, which includes the ultrastatic class
$y=\mathrm{const}$ of zero tidal force. The denominator of (\ref{slope}) is
then positive and flare-out requires
\begin{equation}
\alpha + \beta\sqrt{x_0} \;<\; 0
\quad\Longleftrightarrow\quad
r_0 \;>\; \frac{2\alpha}{|\gamma|}\qquad (\gamma<0).
\label{floor}
\end{equation}
In GR ($\alpha=1$, $\beta=0$) the left inequality is unsatisfiable: the
ultrastatic isotropic class is empty, and more generally no isotropic
wormhole with non-decreasing redshift exists. In Vegh--dRGT gravity the
class exists, switched on by the graviton coupling and only for the sign $\gamma<0$. But
(\ref{floor}), however, has an important consequence. For $\alpha>0$,
$\gamma=-|\gamma|<0$ the zeros of the vacuum
$Z_{\rm vac} = \alpha + \tfrac{1}{2}\gamma r - 2M/r$ solve
$-\tfrac{1}{2}|\gamma|r^2 + \alpha r - 2M = 0$, giving
\begin{equation}
r_\pm = \frac{\alpha \pm \sqrt{\alpha^2 - 4|\gamma|M}}{|\gamma|},
\label{rc}
\end{equation}
which are real only for $0\leq M\leq \alpha^2/(4|\gamma|)$. Two cases
arise. For $0\leq M\leq \alpha^2/(4|\gamma|)$ the vacuum has a static patch
bounded by the outer zero $r_c\equiv r_+\leq 2\alpha/|\gamma|$, and
(\ref{floor}) places the throat beyond it, $r_0>2\alpha/|\gamma|\geq r_c$;
for $M>\alpha^2/(4|\gamma|)$ the discriminant is negative and the vacuum
has no static region at all. In either case the non-decreasing-redshift
branch admits no static-vacuum completion. Such a wormhole cannot be
joined to any static region of its own vacuum; it is a super-horizon
structure. We arrive at a dichotomy for isotropic wormholes in this
theory: either the redshift function is non-decreasing at the throat, and
the wormhole exists only beyond the cosmological zero of the vacuum, with
no static exterior completion; or the wormhole is sub-horizon, in which
case (\ref{slope}) forces
\begin{equation}
2x_0\,\dot y(x_0) + y(x_0) \;<\; 0,
\label{steep}
\end{equation}
equivalently $\bigl[d\ln y/d\ln x\bigr]_{x_0} < -\tfrac{1}{2}$: a
redshift function falling faster than $x^{-1/2}$ at the throat.

We emphasise the logical order: it is the horizon obstruction itself ---
a purely geometric consequence of the slope formula --- that excludes any
static-vacuum completion of the non-decreasing-redshift branch. The compact-star analysis of this theory bounds the coupling,
$|\gamma_1| = |m^2c_1C| \lesssim 4\times10^{-2}\,\mathrm{km}^{-1}$, so for
$\alpha\approx 1$ the excluded branch is additionally confined to
$r_0 \gtrsim 50\,\mathrm{km}$. Static isotropic wormhole candidates admitting a static Vegh-vacuum
exterior must belong to the steep-redshift branch (\ref{steep}). We now exhibit exact solutions realising each branch.

\section{The ultrastatic wormhole: an exact super-horizon solution}
\label{sec:ultra}

\begin{figure}[t]
\includegraphics[width=\columnwidth]{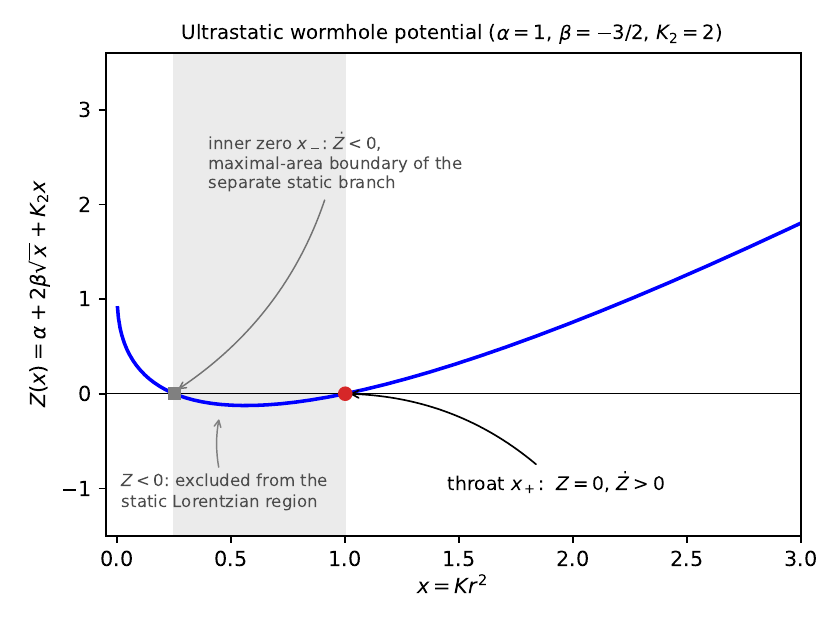}
\caption{The ultrastatic potential (\ref{Zu}) for $\alpha=1$,
$\beta=-3/2$, $K_2=2$. The outer zero is the throat; the inner zero,
where $\dot Z<0$, is the maximal-area boundary of the separate static
branch $x\leq x_-$ and plays no role in the wormhole domain $x\geq x_+$;
the shaded interval has $Z<0$ and is excluded from the static Lorentzian
region.}
\label{f:upot}
\end{figure}

\begin{figure}[t]
\includegraphics[width=\columnwidth]{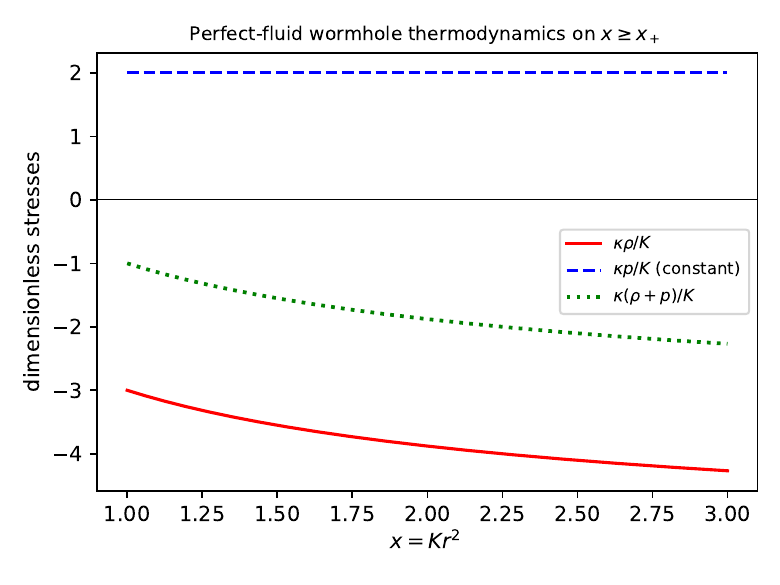}
\caption{Density, uniform pressure, and the radial NEC combination of the
ultrastatic wormhole on $x\geq x_0$. The violation $\rho+p<0$ is carried
entirely by the matter, as the transparency theorem requires.}
\label{f:uthermo}
\end{figure}

\begin{figure}[t]
\includegraphics[width=\columnwidth]{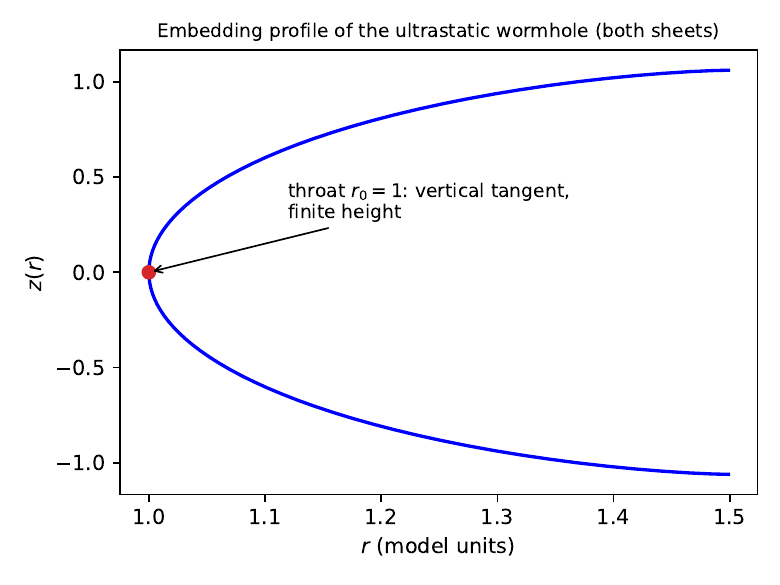}
\caption{Embedding profile of the representative ultrastatic wormhole
(both sheets), on the region $Z<1$.}
\label{f:uembed}
\end{figure}

The ultrastatic condition $y=\mathrm{const}$ annihilates the derivative
terms of (\ref{master}), which then demands precisely the reduction whose general solution is
\begin{equation}
Z(x) = \alpha + 2\beta\sqrt{x} + K_2\,x .
\label{Zu}
\end{equation}
The stellar interpretation requires $Z>0$ on a shell;
the wormhole interpretation selects the complementary regime
$\beta^2 > \alpha K_2$, $\beta<0$, $K_2>0$, in which $Z$ has two positive
zeros $\sqrt{x_\pm} = \bigl[|\beta|\pm\sqrt{\beta^2-\alpha K_2}\bigr]/K_2$.
Writing $u_\pm=\sqrt{x_\pm}$ and $D=\beta^2-\alpha K_2>0$, the slopes at
the two zeros take the compact form
\begin{equation}
\dot Z(x_+) = +\frac{\sqrt{D}}{u_+} \;>\;0,
\qquad
\dot Z(x_-) = -\frac{\sqrt{D}}{u_-} \;<\;0,
\end{equation}
so the outer zero is a throat, with flare-out equivalent to
$|\beta|\sqrt{x_+}>\alpha$, reproducing (\ref{floor}). The inner zero is
\emph{not} a horizon --- $g_{tt}=-1$ is finite and nonzero there --- but a
maximal-area two-sphere bounding the disjoint Lorentzian branch
$x\leq x_-$; the interval between the roots has $Z<0$ and is excluded, and
the wormhole occupies $x\geq x_+$. The matter is a genuine
perfect fluid with
\begin{equation}
\frac{\kappa\rho}{K} = -3K_2 - \frac{2\beta}{\sqrt x},\qquad
\frac{\kappa p}{K} = K_2 = \mathrm{const},
\label{uthermo}
\end{equation}
so the pressure is uniform and positive while the density is negative on
the wormhole domain --- the fluid is exotic through its density, and
$\rho+p<0$ everywhere, consistently with the transparency theorem. The
representative case $\alpha=1$, $\beta=-3/2$, $K_2=2$, with throat at
$x_0=1$, $\dot Z(x_0)=1/2$, $\kappa\rho/K=-3$ and $\kappa p/K=2$, is shown
in Figs.~\ref{f:upot}--\ref{f:uembed}. Asymptotically
$\kappa\rho/K\to-3K_2$ and $\kappa p/K = K_2$, an equation of state
$w=-1/3$: the wormhole is supported at large distance by a curvature-like
halo, and the geometry is not asymptotically flat --- as it cannot be,
since by the dichotomy this solution lives beyond the cosmological zero of
the vacuum and admits no static vacuum completion. It stands as the
explicit and, to our knowledge, first exact demonstration that the
ultrastatic isotropic class, empty in GR, is populated in massive gravity,
while simultaneously illustrating why that population is incompatible with
a static astrophysical exterior.

\section{A sub-horizon exact family from a steep seed}
\label{sec:sub}

The steep-redshift branch (\ref{steep}) is realised most simply by the
seed
\begin{equation}
y(x) = \frac{1}{x}, \qquad e^{2\nu} \propto \frac{1}{x^{2}},
\label{seed}
\end{equation}
for which $2x\dot y+y = -1/x < 0$ everywhere. Substituting (\ref{seed})
into (\ref{linearZ}) gives the linear equation
$x\dot Z - 7Z = \alpha+\beta\sqrt{x}$, whose general solution is
\begin{equation}
Z(x) = c_7\,x^{7} - \frac{\alpha}{7} - \frac{2\beta}{13}\sqrt{x},
\label{Zsub}
\end{equation}
verified against the master equation symbolically. The integration
constant tunes the throat: imposing $Z(x_0)=0$ fixes
$c_7 = \bigl[\alpha/7 + (2\beta/13)\sqrt{x_0}\bigr]/x_0^{7}$ and yields the
remarkably compact throat data
\begin{equation}
\begin{aligned}
\dot Z(x_0) &= \frac{\alpha+\beta\sqrt{x_0}}{x_0}, \qquad
\frac{\kappa\rho}{K}\bigg|_{x_0} &= -\frac{\alpha}{x_0}, \\[4pt]
\frac{\kappa p}{K}\bigg|_{x_0} &= \frac{2|\beta|\sqrt{x_0}-\alpha}{x_0}.
\end{aligned}
\label{throatdata}
\end{equation}
Three features deserve emphasis. First, flare-out now requires
$\alpha+\beta\sqrt{x_0}>0$, i.e.\ $\sqrt{x_0} < \alpha/|\beta|$: the throat
sits \emph{inside} the cosmological radius, so this family is the
astrophysically relevant branch of the dichotomy. Second, the throat density takes the strikingly simple value
$\rho(x_0) = -\alpha K/(\kappa x_0)$, carrying no trace of the graviton
coupling $\beta$; as the seed tower below makes explicit through
(\ref{rhotower}), this value is particular to the $s=1$ family, in
contrast with the throat pressure of the third feature, which is
seed-independent at every isotropic throat. The matter is unavoidably
exotic at the throat, as the theorem demands. Third, the throat
\emph{pressure} changes sign at $\sqrt{x_0} = \alpha/(2|\beta|)$: on the
graviton-generated window
\begin{equation}
\frac{\alpha}{2|\beta|} \;<\; \sqrt{x_0} \;<\; \frac{\alpha}{|\beta|},
\label{window}
\end{equation}
the throat is held by exotic density with \emph{positive} pressure, whereas
in the GR limit $\beta\to 0$ the window closes and the throat can only be a
tension, $p(x_0) = -\alpha K/(\kappa x_0) < 0$. The graviton term thus
converts throat tension into ordinary pressure --- a distinct qualitative
effect of the massive sector, complementary to the transparency theorem. Figures~\ref{f:subZ} and \ref{f:subthermo} display
the representative case $\alpha=1$, $\beta=-3/2$, $\sqrt{x_0}=1/2$, which
sits squarely in the window (\ref{window}): the throat lies inside the
static patch of the vacuum, and a finite junction window separates it from
the cosmological zero.

In fact the throat pressure in (\ref{throatdata}) is far more general than
the seed that produced it. Evaluating the second of (\ref{thermo}) on any
throat kills the $\dot y/y$ term along with $Z$, so that
\begin{equation}
\frac{\kappa p}{K}\Big|_{x_0} = \frac{2|\beta|\sqrt{x_0}-\alpha}{x_0}
\qquad\text{for \emph{every} isotropic wormhole,}
\label{universalp}
\end{equation}
independently of the redshift profile: the window (\ref{window}) of
positive throat pressure is universal, and its lower edge
$\sqrt{x_0}=\alpha/(2|\beta|)$ is a seed-independent landmark of the
theory. The seed dependence resides entirely in the slope and hence in the
throat density. Indeed the single seed (\ref{seed}) generalises to the
tower $y = x^{-s}$ with $s>1/2$, for which (\ref{linearZ}) integrates to
\begin{equation}
Z_s(x) = c\,x^{\frac{4s^2+4s-1}{2s-1}}
- \frac{\alpha}{4s^2+4s-1}
- \frac{2\beta\sqrt{x}}{8s^2+6s-1},
\label{tower}
\end{equation}
verified symbolically for general $s$, with throat slope
$\dot Z(x_0) = (\alpha+\beta\sqrt{x_0})/[(2s-1)x_0]$ --- sub-horizon
flare-out for the entire tower --- and throat density
\begin{equation}
\frac{\kappa\rho}{K}\Big|_{x_0}
= \frac{\alpha(2s-3) - 4|\beta|\sqrt{x_0}\,(s-1)}{(2s-1)\,x_0}.
\label{rhotower}
\end{equation}
Writing $w \equiv |\beta|\sqrt{x_0}/\alpha \in (0,1)$, equation
(\ref{rhotower}) is positive below the curve $w = (2s-3)/[4(s-1)]$, which
exists for $s>3/2$ and asymptotes to the universal pressure landmark
$w=1/2$ as $s\to\infty$. The $(s,w)$ plane thus splits into a phase
diagram, Fig.~\ref{f:seedmap}: throats with positive pressure and negative
density ($w>1/2$), throats with positive density held by tension
($s>3/2$, $w$ small), and a wedge in which both are negative. Positive
density and positive pressure never coexist at a throat --- an immediate
consequence of the transparency identity, since
$\rho+p = -2\dot Z(x_0)K/\kappa < 0$ there --- so the diagram exhausts the
possibilities, and steep seeds trade throat exoticity between the density
and pressure channels in a completely explicit way.

\subsection{Role of the seed}
\label{ssec:seed}

Because the redshift function is the free input of this scheme, it is worth stating precisely which of the foregoing conclusions can be moved by changing the seed. Three cannot. The transparency identity (\ref{transparency}) was established for an arbitrary static spherically symmetric metric, with no field equation imposed, and therefore holds for every seed. Its corollary (\ref{necthroat}) is likewise an identity at
any throat: whichever $y$ is chosen, the matter violates the radial null energy condition there, and the magnitude of the violation is fixed by $b'(r_{0})$ alone. The throat pressure (\ref{universalp}) is seed-independent for the same reason, so the window (\ref{window}) and
its lower edge $\sqrt{x_{0}}=\alpha/(2|\beta|)$ are properties of the theory rather than of any particular model. The dichotomy of Sec.~\ref{sec:dichotomy} is seed-independent in a slightly different sense: it is exhaustive. Every seed falls on one or the other side of the sign of $2x_{0}\dot y(x_{0})+y(x_{0})$ in (\ref{slope}), and there is no third possibility, so no choice of $y$ can deliver a sub-horizon throat with non-decreasing redshift, nor a super-horizon one with a static
exterior. The conclusions on throat location are fixed by the same identity, and the coupling bound that confines the excluded branch to $r_{0}\gtrsim50$~km is inherited from the compact-star analysis, so it too is independent of any choice of seed.

\begin{figure}[t]
\includegraphics[width=\columnwidth]{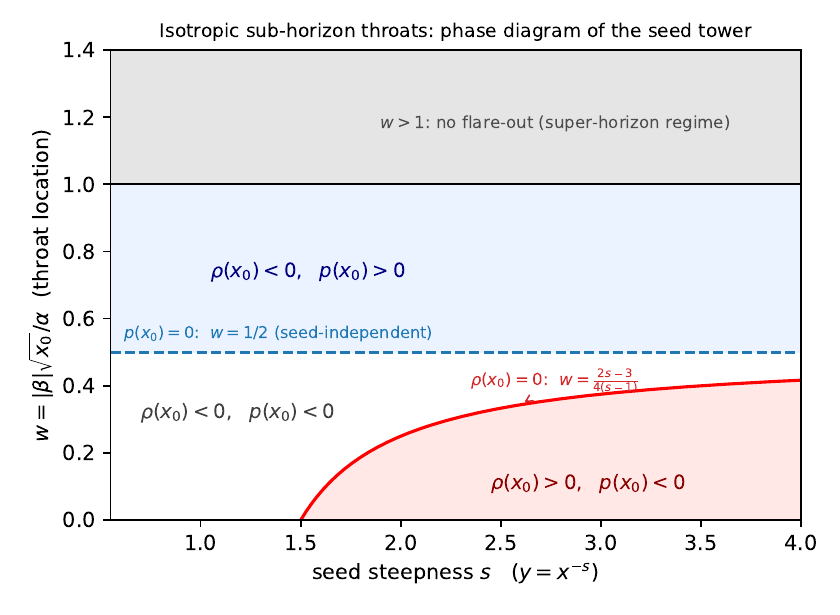}
\caption{Phase diagram of isotropic sub-horizon throats for the seed
tower $y=x^{-s}$, in the plane of steepness $s$ versus throat location
$w=|\beta|\sqrt{x_0}/\alpha$. Flare-out requires $w<1$; the throat
pressure changes sign on the universal, seed-independent line $w=1/2$;
the throat density changes sign on $w=(2s-3)/[4(s-1)]$. Positive density
and positive pressure never coexist, as radial transparency demands.}
\label{f:seedmap}
\end{figure}

\begin{figure}[t]
\includegraphics[width=\columnwidth]{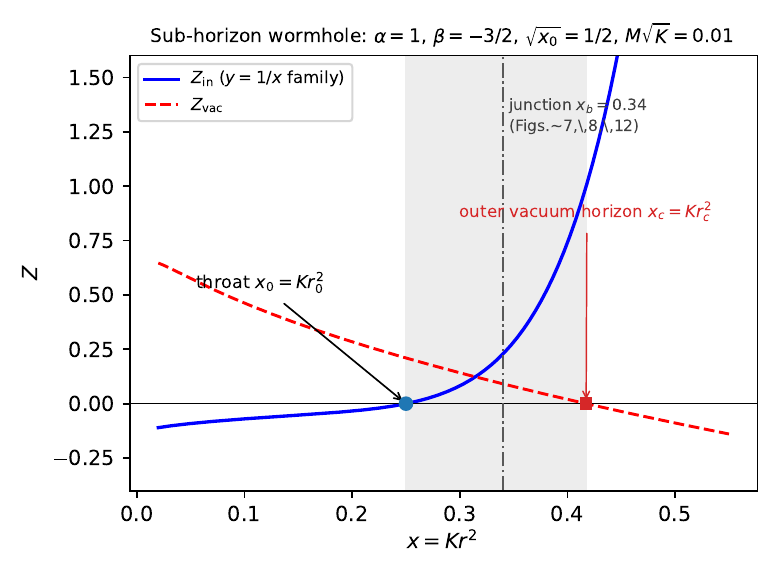}
\caption{The sub-horizon family (\ref{Zsub}) (solid) and the exact vacuum
$Z_{\rm vac}$ (dashed) for $\alpha=1$, $\beta=-3/2$, $\sqrt{x_0}=1/2$ and
$M\sqrt{K}=0.01$. The throat $x_0=Kr_0^2$ lies inside the static patch;
the shaded region is the junction window, bounded by the outer vacuum
horizon $x_c=Kr_c^2$. The dot-dashed line marks the representative
junction $x_b=0.34$ used in Figs.~\ref{f:composite},
\ref{f:composite3D} and \ref{f:deflection}.}
\label{f:subZ}
\end{figure}

\begin{figure}[t]
\includegraphics[width=\columnwidth]{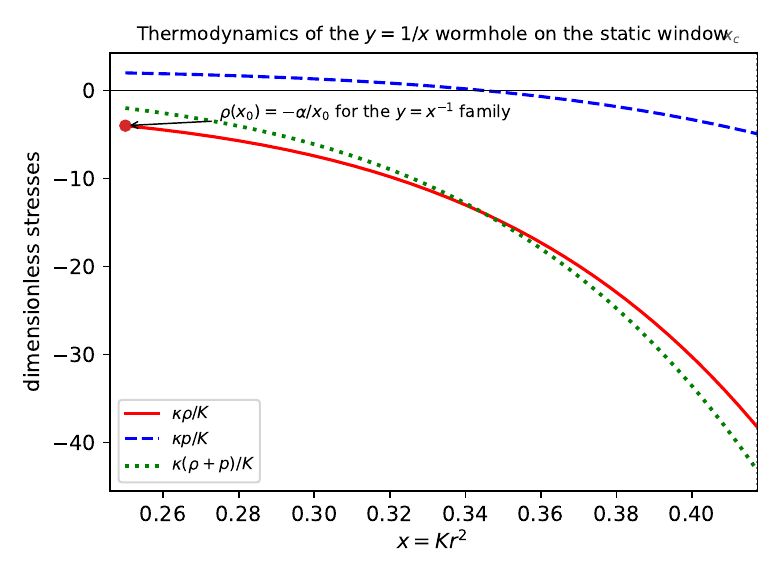}
\caption{Thermodynamics of the $y=1/x$ wormhole, displayed on the static
window $x_0\leq x\leq x_c$. The throat density takes the value
$-\alpha/x_0$, coupling-independent within the $y=x^{-1}$ family, while
the throat pressure is positive on the window (\ref{window}).}
\label{f:subthermo}
\end{figure}

What the seed does control is the slope (\ref{slope}), and through it the throat density, the locations $x_{0}$ at which flare-out can be arranged, the profiles of $\rho$ and $p$ away from the throat, and with them the junction data and the exotic budget. The tower (\ref{tower})  quantifies exactly this freedom within the sub-horizon branch: the throat density (\ref{rhotower}) takes both signs as $s$ and $w$ vary, while Fig.~\ref{f:seedmap} shows the exoticity being traded between the density 
and the pressure channels with $\rho+p$ negative throughout. A different seed redistributes the exotic matter; it does not remove it.

This also settles whether some seed might reproduce the behaviour attributed to Fig.~5 of Ref.~\cite{dutta}. It cannot, and the reason is structural rather than numerical. There the shape function is prescribed 
and flare-out is examined afterwards; here $Z$ is solved for from the seed, and flare-out enters beforehand, as the existence test (\ref{slope}) which a solution must pass before it is called a wormhole at all. A seed returning $\dot Z(x_{0})<0$ is not discarded as unphysical --- it describes a maximal-area sphere, as at $x_{-}$ in
Sec.~\ref{sec:ultra} --- but neither is it admitted as a throat. It is this type of accounting, and not any difference of theory, that separates the two readings of the published numerics.

\section{Darmois--Israel completion and the exotic budget}
\label{sec:shell}

\begin{figure}[t]
\includegraphics[width=\columnwidth]{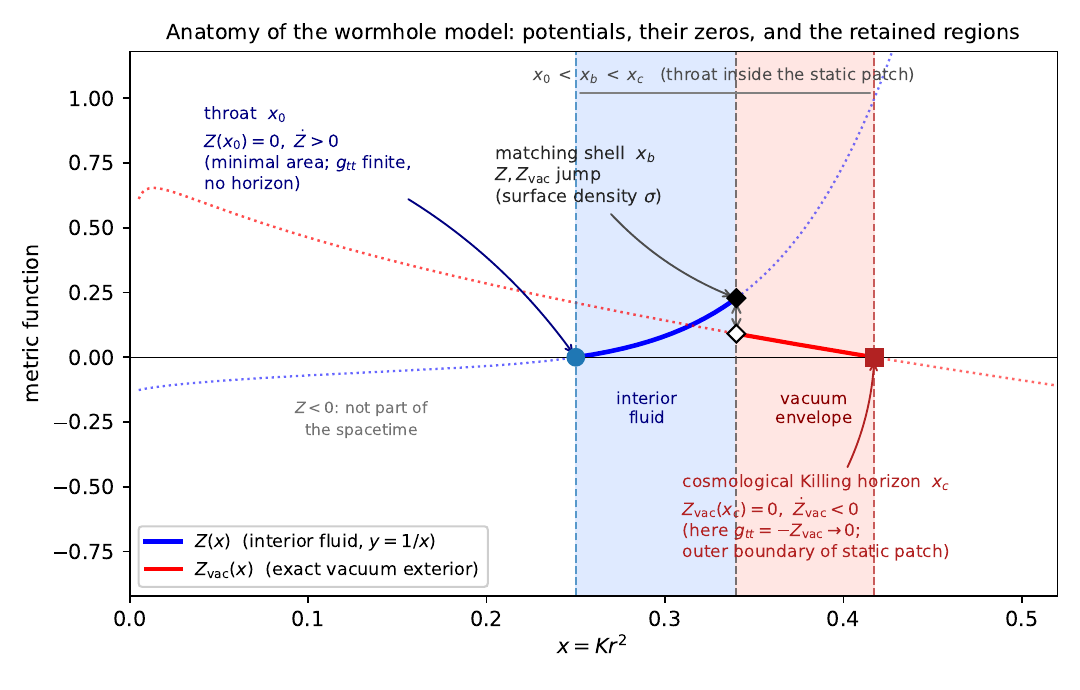}
\caption{How the wormhole model fits together. The interior potential
$Z(x)$ (blue, solid only on the interior fluid interval
$x_0\leq x\leq x_b$) and the exact vacuum potential $Z_{\rm vac}(x)$ (red, solid only on the vacuum interval $x_b\leq x\leq x_c$) are shown together against $x=Kr^2$. Each potential is physical on one side of the shell
only; the dotted continuations are the unphysical extrapolations, retained in neither the geometry nor the field equations. In particular $Z$ and $Z_{\rm vac}$ are never simultaneously active: the vacuum envelope is
governed by $Z_{\rm vac}$ alone, the dotted blue curve there being a mere extrapolation. The final spacetime consists of the interior fluid region $x_0\leq x\leq x_b$ (blue band) joined at the thin shell $x_b$ to the vacuum envelope $x_b\leq x\leq x_c$ (red band). The three loci are
geometrically distinct: the throat $x_0$ is a simple zero of $Z$ with $\dot Z(x_0)>0$, a minimal-area bridge; the matching surface $x_b$ is \emph{not} a zero of either potential but the radius at which $Z$ and $Z_{\rm vac}$ are junction-matched, carrying the surface density $\sigma$
(the double arrow marks the jump); and the cosmological Killing horizon $x_c$ is a simple zero of $Z_{\rm vac}$ with $\dot Z_{\rm vac}(x_c)<0$,
the outer boundary of the static patch. The two zeros differ in kind: at the throat the zero is in the spatial potential, so $g_{rr}=1/Z$ diverges while $g_{tt}=-e^{2\nu}$ stays finite --- a minimal-area bridge, not a
horizon --- whereas at $x_c$ the static vacuum has $g_{tt}=-Z_{\rm vac}$, so its zero is a genuine Killing horizon at which $\partial_t$ turns null.
The ordering $x_0<x_b<x_c$ places the throat strictly inside the static patch. The interval $x<x_0$, where $Z<0$, is not part of the spacetime.}
\label{f:anatomy}
\end{figure}

\paragraph*{The wormhole model assembled.}
The global construction can now be stated explicitly, and is illustrated
in Fig.~\ref{f:anatomy}. The interior potential is the exact isotropic
solution (\ref{Zsub}) generated by the seed $y=1/x$; the exterior is the
vacuum (\ref{Zvac}). Working outward from the throat, the final spacetime
retains three ingredients: the \emph{interior fluid region}
$x_{0}\leq x\leq x_{b}$, on which $Z$ is the physical potential and
$\dot Z>0$; the \emph{thin matching shell} at $x=x_{b}$, where the
interior is joined to the vacuum and the discontinuity in the extrinsic
curvature is supported by the surface stresses $(\sigma,\mathcal{P})$
computed below; and the \emph{vacuum envelope} $x_{b}\leq x\leq x_{c}$, on
which $Z_{\rm vac}$ is the physical potential. Nothing beyond $x_{c}$ is
included, the $t=\mathrm{const}$ slice ceasing to be spatial there, and
extension through that horizon is a separate question. The construction is
then doubled across the throat in the usual Morris--Thorne fashion to give
the two sheets of the bridge. For the representative configuration
($\alpha=1$, $\beta=-3/2$, $\sqrt{x_{0}}=1/2$, $M\sqrt{K}=0.01$) the
ordering (\ref{ordering}) reads $x_{0}=0.25<x_{b}=0.34<x_{c}=0.417$ in
model units, or $r_{0}=0.50<r_{b}=0.58<r_{c}=0.65$.

Because $\rho+p = -(2\dot Z + 4Z/x)K/\kappa$ for the family (\ref{Zsub})
remains negative on its domain, the configuration must be terminated at a
finite junction $x_b$ inside the static patch and joined to the vacuum
$Z_{\rm vac}(x) = \alpha + \beta\sqrt{x} - 2M\sqrt{K}/\sqrt{x}$. Continuity
of the induced metric is arranged by the usual proper-time normalisation on
the shell $\Sigma$ at areal radius $a = \sqrt{x_b/K}$, and the Lanczos
equations give the surface density and surface pressure
\begin{align}
\sigma &= -\frac{1}{4\pi a}\Bigl(\sqrt{Z_{\rm vac}} - \sqrt{Z_{\rm in}}
\Bigr)\Big|_{a},
\label{sigma}\\
\mathcal{P} &= \frac{1}{8\pi}\left[
\frac{\sqrt{Z_{\rm vac}}-\sqrt{Z_{\rm in}}}{a}
+ \frac{Z_{\rm vac}'}{2\sqrt{Z_{\rm vac}}}
+ \frac{2}{a}\sqrt{Z_{\rm in}}
\right]_{a},
\label{Pshell}
\end{align}
where the prime denotes $d/dr$ and the final term in (\ref{Pshell}) is the
interior extrinsic curvature $K^{\tau}{}_{\tau}$ generated by the steep
redshift $e^{\nu}\propto r^{-2}$.
Figure~\ref{f:shell} shows the stresses across the junction window for the
representative case: the surface density changes sign at
$x_b \simeq 0.314$, so that junctions in the outer part of the window carry
\emph{positive} surface energy, with the shell's own NEC combination
$\sigma+\mathcal{P}$ remaining mildly negative throughout. We emphasise
that this surface NEC violation is a property of the particular junction
chosen, not a consequence of the transparency theorem: the theorem
constrains the matter at the throat and places no restriction on outer
matching shells.

\begin{figure}[t]
\includegraphics[width=\columnwidth]{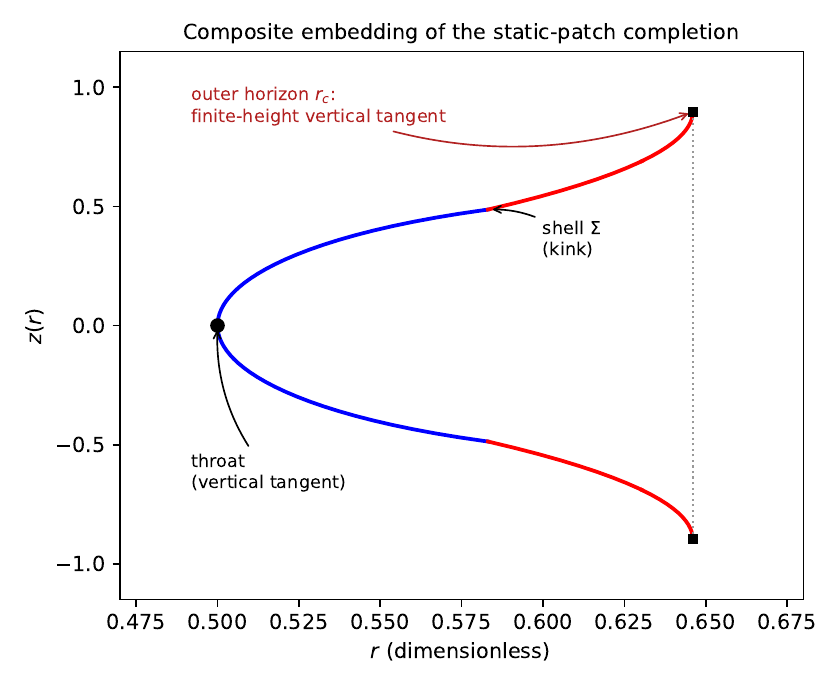}
\caption{Composite embedding of the static-patch completion: interior
$y=1/x$ solution (blue) joined by the shell $\Sigma$ to the exact vacuum
(red). The profile has a vertical tangent at the wormhole throat, a finite
kink at the shell, and a second vertical tangent at the outer vacuum
horizon $r_c$, reached at finite embedding height and finite proper
distance.}
\label{f:composite}
\end{figure}

\begin{figure}[t]
\includegraphics[width=\columnwidth]{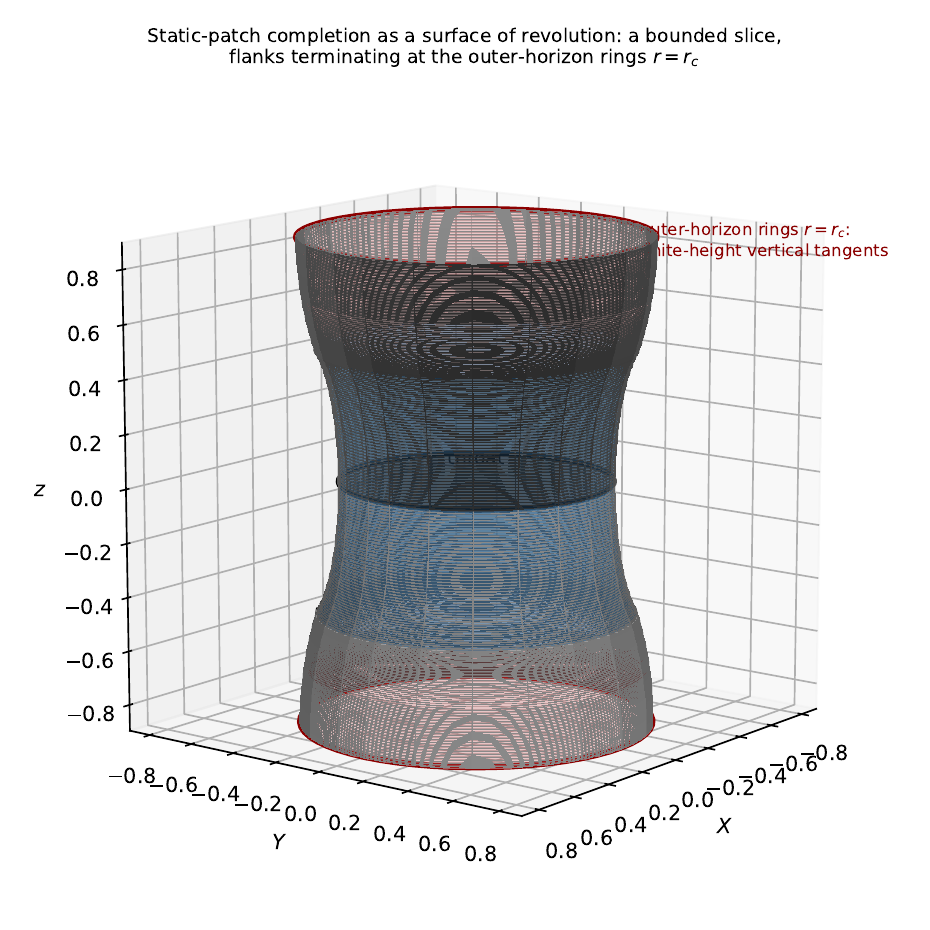}
\caption{The same composite geometry as a surface of revolution: the
traditional hourglass, with the throat ring at the waist, the dashed shell
rings where interior (blue) meets vacuum (red), and the two heavy rings of
radius $r_c$ at which the flanks terminate with vertical tangents on the
outer vacuum horizon. The slice is bounded: it does not open onto flat
planes, nor does it continue as an infinite tube. Axes are equally
scaled.}
\label{f:composite3D}
\end{figure}

\begin{figure}[t]
\includegraphics[width=\columnwidth]{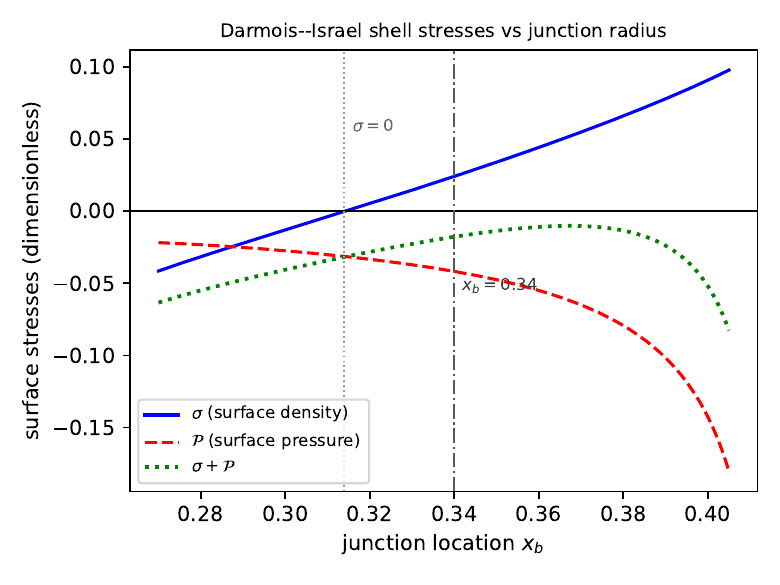}
\caption{Darmois--Israel surface stresses of the completing shell as a
function of the junction location; the dot-dashed line marks the
representative junction $x_b=0.34$. The surface density is positive for
$x_b\gtrsim 0.314$, while for the representative configuration the
combination $\sigma+\mathcal{P}$ remains negative throughout the displayed
interval --- a property of this particular completion, not a requirement
of the transparency theorem.}
\label{f:shell}
\end{figure}

The static-patch completion admits the traditional wormhole portrait.
Because $Z_{\rm in}<1$ on the interior domain and
$Z_{\rm vac} = \alpha + \beta\sqrt{x} - 2M\sqrt{K}/\sqrt{x} < 1$
throughout the static patch (for $\alpha=1$, $\beta<0$, $M>0$), the
composite geometry embeds isometrically in Euclidean $3$-space over the
entire static slice, and both ends of the embedding are governed by the
same local behaviour. Near a simple zero of $Z$ the profile obeys
$(dz/dr)^2 = (1-Z)/Z$, so at the throat
$z-z_0 \simeq \pm 2\sqrt{(r-r_0)/Z_0'}$: a vertical tangent at
\emph{finite} height; and at the outer horizon likewise
$z_c - z \simeq 2\sqrt{(r_c-r)/|Z_c'|}$, with the proper distance
$\ell_c-\ell \simeq 2\sqrt{(r_c-r)/|Z_c'|}$ equally finite.
Figs.~\ref{f:composite} and \ref{f:composite3D} show the familiar flaring
funnel, in profile and as the traditional surface of revolution: a mild
kink at the shell where the extrinsic curvature jumps, and a rim that
\emph{terminates at a vertical tangent on the cylinder} $r=r_c$ --- the
boundary of the static spatial slice --- rather than flattening onto an
asymptotic plane. The correct analogy is the static slice of de Sitter
space, whose embedding is a hemisphere meeting its equator vertically, not
an infinite cylinder; the wormhole flares out normally at its throat, and
the second vertical tangent is caused by the outer Killing horizon of the
Vegh vacuum, not by any failure of flare-out. The $t=\mathrm{const}$ slice
simply ceases to be spatial beyond $r_c$; extension of the spacetime
through the horizon is a separate question not addressed here.

\begin{figure}[t]
\includegraphics[width=\columnwidth]{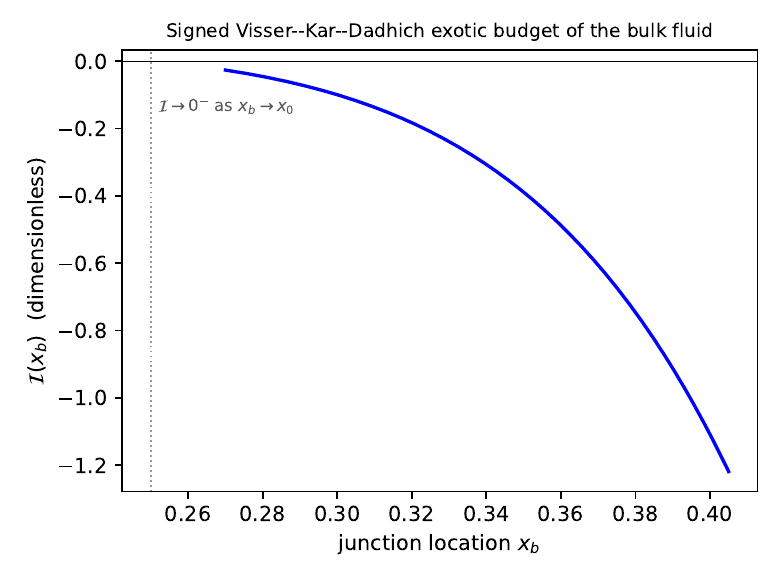}
\caption{The signed Visser--Kar--Dadhich exotic budget (\ref{budget}) of
the bulk fluid across the junction window. It is negative throughout, with
$\mathcal{I}(x_b)\to 0^{-}$ as $x_b\to x_0$, so its magnitude can be made
arbitrarily small by drawing the junction toward the throat.}
\label{f:budget}
\end{figure}

The exotic content of the bulk fluid is finite and explicit. With the
Visser--Kar--Dadhich measure evaluated on the bounded domain
$[x_0,x_b]$ one finds, in closed form,
\begin{equation}
\begin{split}
\mathcal{I}(x_b) &\equiv \int_{x_0}^{x_b}(\rho+p)\,4\pi r^{2}\,dr \\[4pt]
&= -\frac{4\pi K^{-1/2}}{455\,\kappa\,x_0^{7}}
   \Bigl(182\,\alpha\,x_0^{15/2}
   - 260\,\alpha\,x_0^{7}\sqrt{x_b} \\
&\qquad + 78\,\alpha\,x_b^{15/2}
   + 84\,\beta\sqrt{x_0}\,x_b^{15/2} \\
&\qquad + 91\,\beta\,x_0^{8}
   - 175\,\beta\,x_0^{7}x_b\Bigr),
\end{split}
\label{budget}
\end{equation}
verified against direct quadrature to machine precision, and displayed in
Fig.~\ref{f:budget}. The budget vanishes as $x_b\to x_0$ and its
magnitude grows monotonically across the window: as in general
relativity, the exotic bulk content can be made arbitrarily small by
drawing the junction toward the throat \cite{vkd}, and in the present
model the surface stresses of Fig.~\ref{f:shell} remain finite in that
limit --- here realised in exact closed form within massive gravity.

\subsection{Linearised stability of the shell}
\label{ssec:stability}

\begin{figure}[t]
\includegraphics[width=\columnwidth]{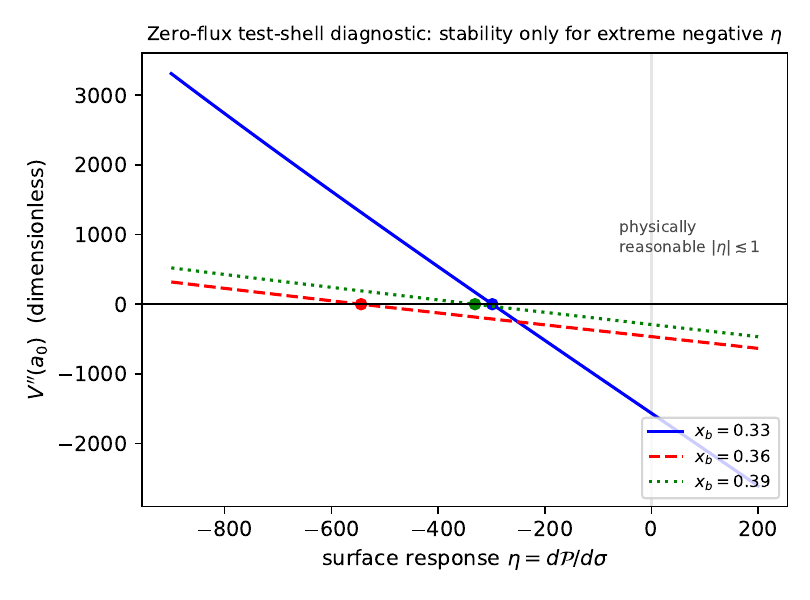}
\caption{Zero-flux test-shell diagnostic: the discriminant $V''(a_0)$ as a
function of the surface response $\eta=d\mathcal{P}/d\sigma$, for three
junction locations $x_b=Ka_0^{2}$ (model units $K=1$). Zero crossings
(dots) occur only near $\eta\simeq-3\times10^{2}$ to $-5\times10^{2}$, far
outside the physically reasonable band $|\eta|\lesssim1$ (shaded), so all
such junctions are radially unstable within this closure.}
\label{f:stability}
\end{figure}

The radial dynamics of the shell follows from promoting the junction to
$a=a(\tau)$; the angular extrinsic curvature becomes
$K^\theta{}_\theta = \sqrt{Z+\dot a^2}/a$ on each side, and the Lanczos
constraint rearranges into the standard energy equation
$\dot a^2 + V(a) = 0$ with
\begin{equation}
V(a) = Z_{\rm in}
- \left[\frac{u}{2} + \frac{Z_{\rm in}-Z_{\rm vac}}{2u}\right]^{2},
\qquad u \equiv 4\pi a\,\sigma(a),
\label{Veff}
\end{equation}
closed by a barotropic surface equation of state with response
$\eta = d\mathcal{P}/d\sigma$ and the test-shell conservation law
$\sigma' = -(2/a)(\sigma+\mathcal{P})$; the flux correction sourced by the
interior fluid is neglected at this first pass and its exact treatment is
deferred. Static solutions are stable iff $V''(a_0)>0$. Evaluating
$V''(a_0)$ across the positive-density branch of the representative
configuration, Fig.~\ref{f:stability}, we find that it depends only weakly
on $\eta$ and vanishes solely near
$\eta\simeq-3\times10^{2}$ to $-5\times10^{2}$, depending on the junction
location: stabilisation would demand a surface response far outside any
physically sensible range. The zero-flux test-shell closure therefore yields radial instability for
the completing shells of the $y=1/x$ configuration; since the interior
fluid has $\rho+p\neq 0$ at the junction, a moving shell generates a
nonvanishing momentum-flux term, and the full physical conclusion awaits
its restoration. As a diagnostic the result is nonetheless in line with
the generic fragility of thin-shell wormholes \cite{poissonvisser}. Shells matched to a vacuum that possesses a cosmological zero are known to be delicate in the same way in general relativity \cite{Lobo:2004linearized}, so the behaviour found here is not peculiar to the mass term. And
the seed tower (\ref{tower}) now allows it to be scanned systematically:
whether steeper seeds, or junctions drawn closer to the throat where the
bulk budget shrinks, admit stable shells is a sharply posed question we
leave to the systematic sweep.

\section{Light bending in the static-patch completion}
\label{sec:deflection}

\begin{figure}[t]
\includegraphics[width=\columnwidth]{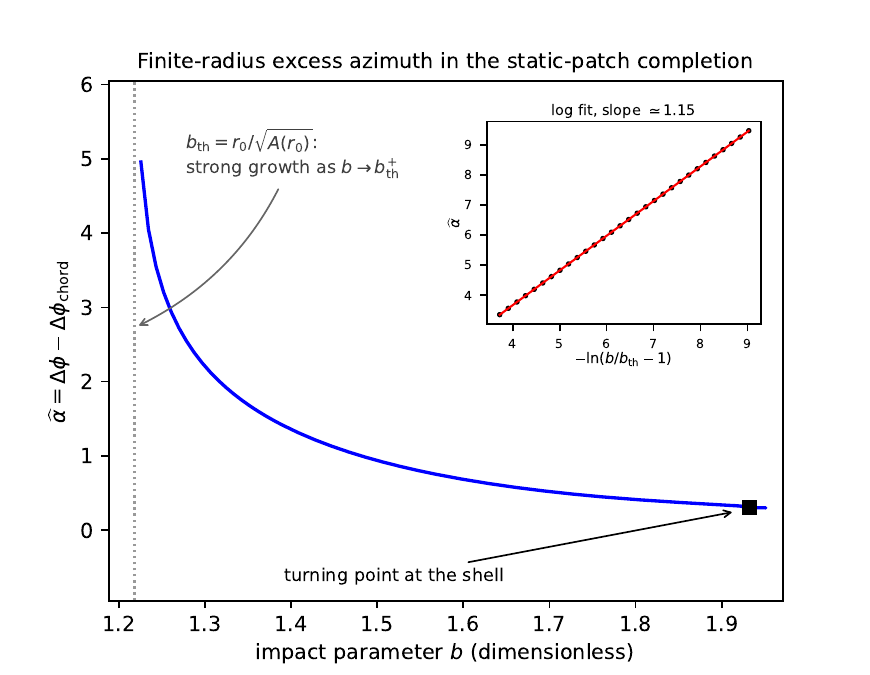}
\caption{Finite-radius excess azimuth
$\Delta\phi-\Delta\phi_{\rm chord}$ of light rays through the static-patch
completion, for an observer at $R_{\rm obs}=0.97\,r_c$. The square marks
turning points at the shell; the dotted line is the throat impact
parameter $b_{\rm th}=r_0/\sqrt{F(r_0)}$, toward which the excess grows
strongly. Inset: the same data against $-\ln(b/b_{\rm th}-1)$, whose
linearity (slope $\simeq1.15$) confirms the logarithmic law expected of an
unstable circular light orbit.}
\label{f:deflection}
\end{figure}

Light bending has already been computed for black holes in this theory, both for particle orbits and lensing in the dRGT background \cite{Panpanich:2019particle} and for the effect of the graviton mass on the deflection angle \cite{Abdusattar:2026investigating}. The wormhole case differs in one respect that controls the whole calculation: the exterior is not asymptotically flat. The completed configuration --- interior, shell, vacuum --- is a concrete
lensing target, and the non-flat asymptotics dictates the observable: the
bending must be referred to emission and reception at finite radii inside
the static patch, rather than to infinity. For a photon with impact
parameter $b = r_{\min}/\sqrt{A(r_{\min})}$, where $A=e^{2\nu}$ is the
(continuous) temporal potential of the composite metric --- written $A$
here to avoid collision with the fiducial profile $F(r)$ of
Sec.~\ref{sec:transp} --- and $H=Z$ its (discontinuous) radial one, the
sweep in azimuth from the turning point to an observer at $R_{\rm obs}$ is
\begin{equation}
\Delta\phi(b) = 2\!\int_{r_{\min}}^{R_{\rm obs}}
\frac{dr}{r^{2}\sqrt{H(r)\bigl[\,b^{-2}A(r)^{-1} - r^{-2}\bigr]}}\,,
\label{deltaphi}
\end{equation}
and we quote the \emph{finite-radius excess azimuth}
$\hat\alpha(b) = \Delta\phi - 2\arccos(r_{\min}/R_{\rm obs})$ relative to
the Euclidean chord between the same radii, which reduces to the textbook
deflection $\Delta\phi-\pi$ when $R_{\rm obs}\to\infty$ in an
asymptotically flat spacetime. In the present non-asymptotically-flat
setting $\hat\alpha$ is a coordinate diagnostic rather than an invariant
deflection angle; the tetrad-based finite-distance observable, which
supplements (\ref{deltaphi}) with the local incidence angles at source and
receiver, is deferred.
Figure~\ref{f:deflection} shows $\hat\alpha(b)$ for the representative
configuration with $R_{\rm obs}=0.97\,r_c$. Rays whose turning points lie
in the vacuum probe the exterior $\alpha+\gamma r/2-2M/r$; as the turning
point crosses the shell the deflection curve is continuous but changes
slope, a lensing fingerprint of the surface layer; and as
$b$ decreases toward the throat value the excess azimuth grows without
bound. The divergence signals an unstable circular photon orbit at the
throat, and this deserves a short proof rather than an appeal to
$H\to 0$: on the reflection-symmetric doubled wormhole the photon motion
is governed by $\mathcal{B}=r^2/A$ along proper radial distance $\ell$;
symmetry forces $d\mathcal{B}/d\ell=0$ at the throat; for the present seed
$A\propto r^{-4}$ gives $\mathcal{B}\propto r^{6}$, and the flare-out
property
$d^2r/d\ell^2>0$ then makes the throat a minimum of $\mathcal{B}$ ---
equivalently a maximum of the photon potential $A/r^2$ --- hence an
unstable circular light orbit with the attendant logarithmic
strong-deflection divergence; the inset of Fig.~\ref{f:deflection}
confirms this numerically, the excess azimuth being linear in
$-\ln(b/b_{\rm th}-1)$ with slope $\simeq1.15$ as $b\to b_{\rm th}^{+}$.
The check is not vacuous:
reflection-symmetric throats can support stable or unstable light rings
depending on the metric. A full strong-deflection-limit expansion of
(\ref{deltaphi}), and the image structure it implies --- with the caveat
that a traversable throat transmits photons between the two universes, so
any shadow depends on the illumination of both sides --- are natural
continuations.

\section{Discussion and outlook}
\label{sec:disc}

The results assemble into a coherent picture. The graviton sector of
Vegh--dRGT massive gravity is transparent to the radial null energy
condition, so it can never substitute for the exotic matter that a
Morris--Thorne throat requires; under their own effective-stress
decompositions, published claims to the contrary are incompatible with the
flare-out identity. What
the graviton sector does instead is legislate \emph{where} throats may
exist and \emph{how} they are dressed: isotropic wormholes with
non-decreasing redshift are pushed beyond the cosmological zero of the
theory's own vacuum --- and, through the compact-star bound of the
companion paper, beyond fifty kilometres --- while astrophysical throats
admitting a static Vegh-vacuum exterior are confined to the
steep-redshift branch, for which we have exhibited
an exact family with a universal, coupling-independent throat density and a
graviton-generated window of positive throat pressure, completed by a thin
shell whose stresses and whose exotic budget are known in closed form.

Several continuations suggest themselves. The seed tower (\ref{tower})
maps the isotropic sub-horizon space explicitly, and its phase diagram
shows the throat trading exoticity between density and pressure channels
around the universal landmark $w=1/2$; extending the linearised shell
analysis across the tower, with the interior flux term restored, will
determine whether stable completions exist. The anisotropic sector with
the shape function derived rather than imposed, the strong-deflection
expansion of (\ref{deltaphi}) with its image ladder and
transmission-dependent shadow structure, and the throat perturbation
problem all lie within reach of the present formalism. Finally, the interplay established here
between compact-star constraints and wormhole existence—one and the
same coupling window governing both—suggests that multimessenger bounds
on the graviton mass translate, in this theory, into exclusion statements
about entire classes of exotic compact objects, a direction we intend to
pursue.



\section*{Declaration of competing interest}
The authors declare that there are no financial or personal relationships that could have influenced the research
presented in this paper.

\section*{ Data availability}

The data supporting the findings of this study are available from the corresponding author upon reasonable request. All numerical results presented in this work were obtained from the theoretical framework and equations described in the manuscript.

\bibliographystyle{apsrev4-2}
\bibliography{references}

\end{document}